\documentclass[preprint,showpacs,preprintnumbers,amsmath,amssymb]{revtex4}
\usepackage{amsmath,amssymb,graphics,epsfig,subfigure}

\begin{document}
\newcommand {\nn}    {\nonumber}
\renewcommand{\baselinestretch}{1.3}

\title{Bulk Matter Fields on Two-field Thick Branes}
\author{Chun-E Fu\footnote{fuche08@lzu.cn},
        Yu-Xiao Liu\footnote{liuyx@lzu.edu.cn, corresponding author},
        Heng Guo\footnote{guoh06@lzu.cn}}.
 \affiliation{Institute of Theoretical Physics,
              Lanzhou University, Lanzhou 730000,
             People's Republic of China}

\begin{abstract}

In this paper we obtain a new solution of a brane made up of a scalar field coupled to a dilaton. There is a unique parameter $b$ in the solution, which decides the distribution of the energy density and will effect the localization of bulk matter fields. For free vector fields, we find that the zero mode can be localized on the brane. And for vector fields coupled with the dilaton via $\text{e}^{\tau\pi}F_{MN}F^{MN}$, the condition for localizing the zero mode is $\tau\geq-\sqrt{b/3}$ with $0<b\leq1$, or $\tau>-1/\sqrt{3b}$ with $b>1$, which includes the case $\tau=0$. While the zero mode for free Kalb-Ramond fields can not be localized on the brane, if only we introduce a coupling between the Kalb-Ramond fields and the dilaton via $\text{e}^{\zeta \pi}H_{MNL}H^{MNL}$. When the coupling constant satisfies $\zeta>1/\sqrt{3b}$ with $b\geq1$ or $\zeta>\frac{2-b}{\sqrt{3b}}$ with $0<b<1$, the zero mode for the KR fields can be localized on the brane. For spin half fermion fields, we consider the coupling $\eta\bar{\Psi}\text{e}^{\lambda \pi}\phi\Psi$ between the fermions and the background scalars with positive Yukawa coupling $\eta$. The effective potentials for both chiral fermions have three types of shapes decided by the relation between the dilaton-fermion coupling constant $\lambda$ and the parameter $b$. For $\lambda\leq-1/\sqrt{3b}$, the zero mode of left-chiral fermion can be localized on the brane. While for $\lambda>-1/\sqrt{3b}$ with $b>1$ or $-1/\sqrt{3b}<\lambda<-\sqrt{b/3}$ with $0<b\leq1$, the zero mode for left-chiral fermion also can be localized.

\end{abstract}




\maketitle

\section{Introduction}

The brane theory \cite{Rubakov1983,Akama}, which considers our four-dimensional universe as a hyper-surface (``brane world") embedded in more higher dimensional space-time, has received a great of renewed attention. In this theory all matter fields are confined to the brane in a high-dimensional space, while only gravity is free to propagate in all dimensions. The extra dimensions can be compact \cite{Antoniadis,ADD,RS1}, or infinite and non-compact \cite{RS2,GRS}. This theory has opened up new avenues to explain some questions in particle physics and in astrophysics, such as the hierarchy problem, the cosmological problem, the nature of
dark matter and dark energy
\cite{ADD,RS1,RS2,Gogberashvili,apply1,apply2,braneDarkEnergy}.

Some brane models consider the brane as infinitely thin branes with delta-like localization of matter, which are ideal models \cite{ADD, RS1,RS2}. So some thick brane models are proposed. The thick branes are usually realized naturally by one \cite{DeWolfe,GremmPLB2000,ThickBrane1,ThickBraneWeyl,varios,1006.4240,
Guerrero2002,ThickBraneDzhunushaliev,ThickBraneBazeia,ShtanovJCAP2009,
Bazeia0808,guo10} or two \cite{two-scalarbrane1,BlochBrane,BranewithTwoScalars}
background scalar fields configuration coupled with gravity. For a comprehensive review on thick brane solutions and related topics please see Ref.~\cite{Dzhunushaliev}.

In this paper, we will investigate the localization of various matter fields on a thick brane generalized by two background scalar fields, i.e., a kink scalar and a dilaton scalar, which is similar to that in
Refs.~\cite{two-scalarbrane1,KRthickbrane}. But in our solution of the brane there is a unique parameter $b$, which makes the solution in Refs.~\cite{two-scalarbrane1,KRthickbrane} only one case of us. The unique parameter $b$ decides the distribution of the energy density of the bulk, and will effect the localization of various bulk matter fields differently.

The localization of various matter fields on the branes is an important problem in the braneworld theory, which in order to build up the standard model. It has been known that massless scalar fields and graviton can be localized on branes of different types \cite{RS1,RS2,BajcPLB2000} with exponentially decreasing warp factor. But spin-1 Abelian vector fields only can be localized on the RS brane in some higher-dimensional cases \cite{OdaPLB2000113}, or on the thick de Sitter brane and the Weyl thick brane \cite{LiuJCAP2009,Liu0803}.

The anti-symmetric Kalb-Ramond (KR) tensor field $B_{\mu\nu}$ was first introduced in the string theory, in which it is associated with massless modes. Then it was used to explain the torsion of the space-time in the Einstein-Cartan theory. Moreover in the four-dimensional space-time, by a symmetry known as duality, antisymmetric tensor fields are just equivalent to scalar or vector fields \cite{duality}. But in extra dimensions they will indicate new types of particles. Thus any observational effect involving the KR fields is a window into the inaccessible world of very high energy physics. The investigation of the KR fields in the context of theories with extra dimensions has been carried out in Refs.~
\cite{KRfieldinRS1,KRfieldinRS2,KRfieldinRS3,KRthickbrane,KR1006.1366,q-formField}.

In Refs.~\cite{KRfieldinRS1, KRfieldinRS2,KRfieldinRS3}, the authors proved that in the background of RS space-time both the massless and the massive KK modes of the KR fields appear much weaker than curvature to an observer on the visible RS brane, however when the KR fields couple with the dilaton fields, the trilinear dilaton-KR couplings may lead to new signals in Tev scale experiments. While in Ref.~\cite{KRthickbrane}, the author also proved that only when the KR fields couple with the dilaon field, the zero mode of the KR fields can be localized on a thick
brane. But there is only a zero mode and no bound massive KK mode. In our work we also find that it is necessary to introduce the dilaton-KR coupling in order to obtain a localized zero mode, and we find there are massive bound KK modes.

And the localization of the spin $1/2$ fermion fields is also interesting. It has been proved that in order to have normalizable zero modes the fermion fields should couple with the background scalars. With different scalar-fermion coupling, there
may exist a single bound state and a continuous gapless spectrum of massive fermion KK states \cite{1006.4240,Liu0708,Liu0709,20082009,Fermions2010}, or exist finite
discrete KK states (mass gap) and a continuous gapless spectrum starting at a positive $m^2$ \cite{Liu0803,zhaoliu2010,fuGRS,0803.1458}, or even only exist bound KK modes \cite{Liu0907.0910,Liu6}. While in this paper, we will show that with one scalar-fermion coupling to both background scalars, there will exist the above three cases with different relation between the unique parameter $b$ and the dilaton-fermion coupling constant.

Our paper is organized as follows: In Sec. \ref{SecModel}, we give a brief review of the braneworld generated by two scalars. Then, in Sec. \ref{SecLocalize}, we study the localization and mass spectra of the vector, KR, and fermion fields on the brane by presenting the potentials of the corresponding Schr\"{o}dinger equations. Finally, a brief discussion and conclusion are given in the last section.

\section{Review of the brane generated by two interacting scalars}
\label{SecModel}

In this paper we consider the braneworld generated by two interacting scalars $\phi$ and $\pi$. The action of the system is
\begin{equation}\label{action}
 S=\int d^{5}x\sqrt{-g}\bigg[\frac{1}{2\kappa_5^2}R-\frac{1}{2}(\partial\phi)^{2}
   -\frac{1}{2}(\partial\pi)^{2}-V(\phi,\pi)\bigg]
\end{equation}
with $R$ the scalar curvature and $\kappa_5^2=8 \pi G_5$, where $G_5$ is the 5-dimensional Newton constant. Here we set $\kappa_5=1$. The line-element of a 5-dimensional space-time can be assumed as \cite{two-scalarbrane1,KRthickbrane,YZhong2010}
\begin{equation}\label{line-element}
ds^{2}=\text{e}^{2A(y)}\eta_{\mu\nu}dx^{\mu}dx^{\nu}+\text{e}^{2B(y)}dy^{2},
\end{equation}
where $\text{e}^{2A}$ and $\text{e}^{2B}$ are the warp factors and $y$ stands for the extra coordinate. And the background scalars $\phi, \pi$ are assumed to be only the functions of $y$, because the brane can be treated as the cross-section of the bulk. In this model, the thick brane is realized by the potential $V(\phi, \pi)$. Then the equations of motion generated from the action (\ref{action}) with the ansatz (\ref{line-element}) are given by
\begin{eqnarray}\label{EOM1}
 \frac{1}{2}\phi'^2 + \frac{1}{2}\pi'^2 - \text{e}^{2B} V &=& 6A'^2, \\\label{EOM2}
 \frac{1}{2}\phi'^2 + \frac{1}{2}\pi'^2 + \text{e}^{2B} V &=& -6A'^2-3A''+3A'B',\\\label{EOM3}
 \phi'' + (4A' - B')\phi' &=& \text{e}^{2B}\frac{\partial V}{\partial \phi},\\\label{EOM4}
 \pi'' + (4A' - B')\pi' &=& \text{e}^{2B}\frac{\partial V}{\partial \pi},\label{EOM5}
\end{eqnarray}
where the prime stands for the derivative with respect to $y$.

The solutions of the system can be found by following the superpotential method \cite{DeWolfe}. With the superpotential function $W(\phi)$ and suppose $V= \text{e}^{-2\sqrt{b/3}\;\pi}\bigg[\frac{1}{2}(\frac{\partial W}{\partial\phi})^{2}-\frac{4-b}{6}W^{2}\bigg]$, it can be verified that the following first-order differential equations are the solutions of the equations of motion (\ref{EOM1}-\ref{EOM5}):
\begin{eqnarray}
\phi'=\frac{\partial W}{\partial \phi}~~~, A' = -\frac{1}{3}W,~~ B = b A, ~~\pi = \sqrt{3b} A,\label{solution1}
\end{eqnarray}
where $b$ is a positive constant. For a specific superpotential $W(\phi)$
\cite{two-scalarbrane1,KRthickbrane}:
\begin{equation}
W(\phi)=va\phi\left(1-\frac{\phi^{2}}{3v^{2}}\right),
\end{equation}
the solutions are found to be
\begin{eqnarray}
\phi(y)&=&v\tanh(ay),\label{dilat}\\
A(y)&=&-\frac{v^2}{9}\bigg(\ln\cosh^2(ay)
                        +\frac{1}{2}\tanh^2(ay)\bigg), \label{warpfactor}\\
\pi(y)&=&\sqrt{3b} \;A(y),\\
B(y)&=& b \;A(y),
\end{eqnarray}
where $v,a$ are both positive constants. It can be seen that the solution for $\phi$ is a kink and $\pi=\sqrt{3b}A$ is the dilaton field consistent with the metric and the kink. The solutions in Refs.~\cite{two-scalarbrane1,KRthickbrane} are only one case of above solutions when we let $v^2/9=\beta$ and $b=1/4$. And here we have another parameter $b$, which leads to new solutions of the brane world. Our solutions for the brane are not amount to a simple coordinate change, because the solutions are decided by the scalar potential $V(\phi)$ with the parameter $b$, which does not depend on the coordinate systems.

In order to clarify this question more clearly, we would like to discuss the effect of the parameter $b$ on the brane under the physical coordinate $\bar{y}$. To this end, we perform a coordinate transformation $dy=\text{e}^{-b A}d\bar{y}$ to translate the different coordinate $y$ to the same physical coordinate $\bar{y}$. Then the metric is read as
\begin{equation}\label{line-element2}
ds^{2}=\text{e}^{2A(y(\bar{y}))}\eta_{\mu\nu}dx^{\mu}dx^{\nu}+d\bar{y}^{2}.
\end{equation}
With the relation of the two coordinate systems we have
\begin{equation}
\bar{y}=\int_0^{y}\text{e}^{b A}d\tilde{y}\rightarrow
       \int_0^{y}\text{e}^{-\frac{2v^2a}{9}b\tilde{y}}d\tilde{y},
       ~~~~\text{for}~~~\tilde{y}\rightarrow\infty,
\end{equation}
so it can be seen that in this new coordinate system the extra dimension $\bar{y}$
is finite (with $\bar{y}_{max}=\frac{9}{2v^2ab}$), which is different from the un-physical and infinite coordinate $y$. For different $b$, the boundary of the extra dimension $\bar{y}$ is also different.

\begin{figure*}[htb]
\begin{center}
\includegraphics[width=4.5cm]{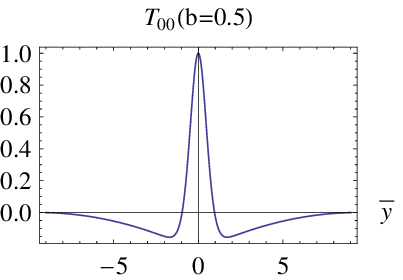}
\includegraphics[width=4.5cm]{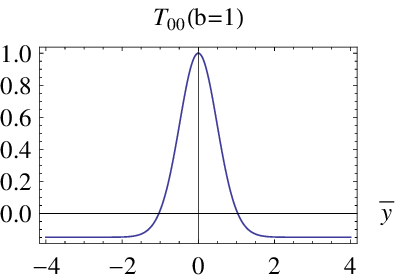}
\includegraphics[width=4.6cm]{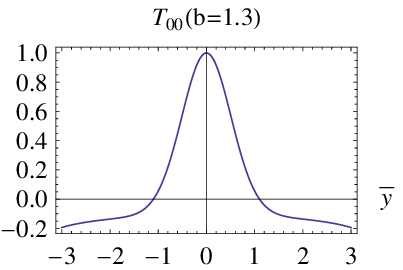}
\includegraphics[width=4.0cm]{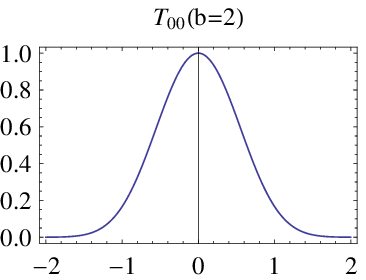}
\includegraphics[width=4.5cm]{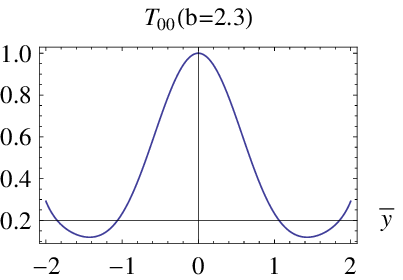}
\end{center}
 \caption{The shapes of the energy density $T_{00}(\bar{y})$ for different $b$. The parameters
are set to $a=1,v=1$.}
 \label{figT00}
\end{figure*}

Now we can investigate the energy density of the system $T_{00}(\bar{y})$ in the new coordinate. With the solution (\ref{warpfactor}) and the relation between the two coordinates we calculate the values of the energy density $T_{00}(\bar{y})$ at $\bar{y}=0$ and at $|\bar{y}|\rightarrow \bar{y}_{max}$:
\begin{eqnarray}
T_{00}(0)&=& a^2v^2,\\
T_{00}(|\bar{y}|\rightarrow\bar{y}_{max})&\rightarrow&\frac{4a^2v^4}{27}(b-2)
\bigg(-\frac{2v^2a\;b}{9}\bar{y}+1\bigg)^{-2(b-1)/b},
\end{eqnarray}
from which we can see that the behavior of $T_{00}(|\bar{y}|\rightarrow\bar{y}_{max})$ at the boundaries of the extra dimension is:
\begin{subequations}
\begin{eqnarray}
 T_{00}(|\bar{y}|\rightarrow\bar{y}_{max}) &=& \left\{
    \begin{array}{ll}
      0, ~~~~~~~~~~~~~~  & 0<b<1    \\
      -\frac{4}{27}a^2v^4, ~~~~~~ &b=1 \\
      -\infty, ~~~~~~~~~~~~ & 1<b<2 \\
      0, ~~~~~~~~~~~~~~~  & b=2    \\
      \infty,~~~~~~~~~~~~~ & b>2
    \end{array}. \right.
\end{eqnarray}
\end{subequations}
We plot the energy density $T_{00}(\bar{y})$ for different $b$ using numerical method in Fig.~\ref{figT00}.

In the following we will mainly discuss the effect of the parameter $b$ on the localization of bulk matter fields.

\section{Localization and mass spectra of various bulk matter fields on the brane}
\label{SecLocalize}

In this section we will investigate the localization and mass spectra of various bulk matter fields in this braneworld by presenting the potentials of the corresponding Schr\"{o}dinger equations. In order to make sure the solutions of the system obtained before valid, we treat the bulk matter fields considered below as perturbations around the background \cite{9808016,9905186}, namely, we neglect the back-reaction of bulk matter fields on the background geometry. And we will use the conformally flat metric
\begin{eqnarray}
 ds^2&=&\text{e}^{2A(z)}\big(\eta_{\mu\nu}dx^\mu dx^\nu
          + dz^2\big). \label{Clinee}
\end{eqnarray}
Comparing it with the metric (\ref{line-element}), we find the two coordinate systems
are connected by $dz=\text{e}^{2(b-1)A}dy$. For the conformally flat space-time, the extra dimension $z$ will be infinite for $0<b\leq1$ and finite (with $|z|\leq z_{max}=\frac{9}{2v^2a(b-1)}$) for $b>1$. From the following calculations we can see that the mass-independent potentials can be obtained conveniently with the conformally flat metric (\ref{Clinee}), and we will mainly investigate the effect of the parameter $b$ on the zero modes and the spectra for various bulk matter fields.

\subsection{Spin-1 vector fields}
First we investigate the localization of the spin-1 vector fields in 5-dimensional space. The action of vector
fields coupled with the dilaton is
\begin{eqnarray}
S_1 = - \frac{1}{4} \int d^5 x
\sqrt{-g}\;\text{e}^{\tau\pi} g^{M R} g^{N S} F_{MN}
F_{RS}, \label{actionVector}
\end{eqnarray}
where the field strength tensor is given by $F_{MN} = \partial_M A_N - \partial_N A_M$ and $\tau$ is the coupling constant between the dilaton and the vector field. The equations of motion can be obtained using the background geometry (\ref{Clinee}):
\begin{eqnarray}
\text{e}^{\tau\pi}\frac{1}{\sqrt{-\hat{g}}}\partial_\nu (\sqrt{-\hat{g}}
      {\hat{g}}^{\nu \rho} \hat{g}^{\mu\lambda}F_{\rho\lambda})
    +{\hat{g}^{\mu\lambda}}\text{e}^{-A}\partial_4
      \left(\text{e}^{(A+\tau\pi)} F_{4\lambda}\right)  &=& 0, \\
 \text{e}^{\tau\pi}\partial_\mu(\sqrt{-\hat{g}}\hat{g}^{\mu\nu}F_{\nu 4})&=& 0.
\end{eqnarray}
Then with the gauge choice $A_4=0$ and the decomposition of the vector field $A_{\mu}(x,z)=\sum_n a^{(n)}_\mu(x)\rho_n(z)\text{e}^{-(1+\sqrt{3b}\;\tau)A/2}$, we find that the KK modes of the vector field satisfy the following Schr\"{o}dinger-like equation:
\begin{eqnarray}
  \left[-\partial^2_z +V_1(z) \right]{\rho}_n(z)=m_n^2
  {\rho}_n(z),  \label{SchEqVector1}
\end{eqnarray}
with $m_n$ the masses of the 4-dimensional vectors, and
the effective potential is
\begin{eqnarray}\label{VV}
V_1(z)=\frac{(1+\sqrt{3b}\;\tau)^2}{4}(\partial_z A)^2
            +\frac{1+\sqrt{3b}\;\tau}{2}\partial_z^2 A.
\end{eqnarray}

Furthermore, providing the orthonormality condition
\begin{eqnarray}
 \int^{z_b}_{-z_b} dz \;\rho_m(z)\rho_n(z)&=&\delta_{mn},
 \label{normalizationConditionVecter}
\end{eqnarray}
we can get the 4-dimensional effective action:
\begin{eqnarray}
 S_1 &=& \sum_{n}\int d^4 x \sqrt{-\hat{g}}~
       \bigg( - \frac{1}{4}\hat{g}^{\mu\alpha} \hat{g}^{\nu\beta}
             f^{(n)}_{\mu\nu}f^{(n)}_{\alpha\beta}
             - \frac{1}{2}m^2_{n} ~\hat{g}^{\mu\nu}
           a^{(n)}_{\mu}a^{(n)}_{\nu}
       \bigg),
\label{actionVector2}
\end{eqnarray}
where $f^{(n)}_{\mu\nu} = \partial_\mu a^{(n)}_\nu -\partial_\nu a^{(n)}_\mu$ is the
4-dimensional field strength tensor.

\begin{figure*}[htb]
\begin{center}
\includegraphics[width=5.0cm]{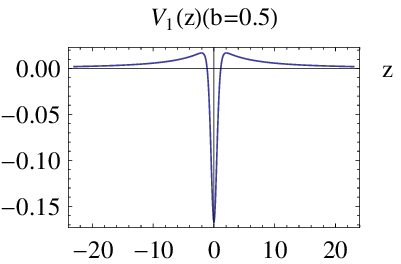}
\includegraphics[width=4.6cm]{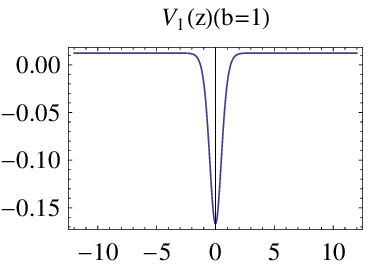}
\includegraphics[width=5.0cm]{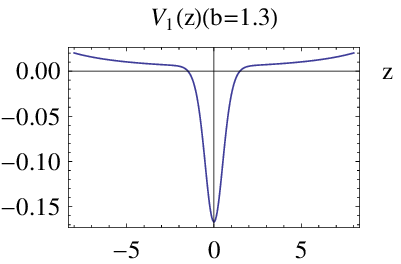}
\includegraphics[width=4.7cm]{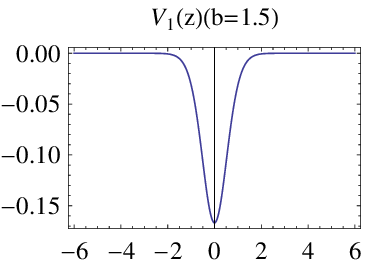}
\includegraphics[width=5.0cm]{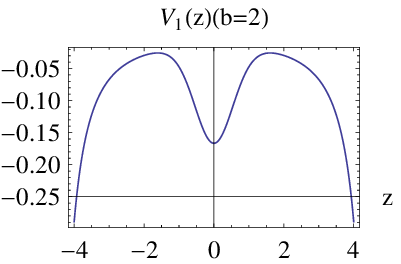}
\end{center}
 \caption{The shapes of the potentials of the vector field $V_1(z)$ for different parameter $b$,
and $a=1,v=1,\tau=0$.}
 \label{figVVector}
\end{figure*}

We rewrite the potential (\ref{VV}) as the function of $y$:
\begin{eqnarray}
V_1(y)=\text{e}^{2(1-b)A}\bigg[\frac{1+\sqrt{3b}\;\tau}{2}\partial_y^2 A
+\frac{(1+\sqrt{3b}\;\tau)^2+2(1+\sqrt{3b}\;\tau)(1-b)}{4}\big(\partial_y A\big)^2\bigg].
\end{eqnarray}
So with the relation $dz=\text{e}^{(b-1)A}dy$, we can get the values of $V_1(z)$ at
$z=0$ and $|z|\rightarrow z_b$:
\begin{eqnarray}
V_1(0)&=&-\frac{1}{6}a^2v^2(1+\sqrt{3b}\;\tau),\\
V_1(|z|\rightarrow z_b)&=&
\frac{v^4a^2(1+\sqrt{3b}\;\tau)(3-2b+\sqrt{3b}\;\tau)}{81[-\frac{2v^2a}{9}(b-1)z+1]^2}.
\end{eqnarray}
In order to get a zero mode, we have to insure that the value of $V_1(z)$ at $z=0$ is negative, the condition is turned out to be
\begin{eqnarray}
\tau>-1/\sqrt{3b},\label{ConditionForTau}
\end{eqnarray}
From which we can see that if $\tau=0$, namely, there is no coupling between the vector and the dilaton field, there could also exist a zero mode. Then with the condition (\ref{ConditionForTau}), the potential $V_1(z)$ is volcano-like and PT-like ones for $0<b<1$ and $b=1$, respectively. And for $b>1$, the potential will divergent at the boundary of the extra dimension with $\tau\neq\frac{2b-3}{\sqrt{3b}}$, but vanish with $\tau=\frac{2b-3}{\sqrt{3b}}$. We plot the shapes of the potential in Fig.~\ref{figVVector}.

Under the condition (\ref{ConditionForTau}), the zero mode for the vector field can be obtained by setting $m_0=0$:
\begin{eqnarray}
\rho_0\varpropto \text{e}^{\frac{(1+\sqrt{3b}\;\tau)}{2}A(z)}.
\end{eqnarray}
In order to check whether the zero mode for the vector field can be localized on the brane, we can investigate whether it satisfies the orthonormality condition $\int^{z_b}_{-z_b} dz \;\rho_m(z)\rho_n(z)=\delta_{mn}$ (or $\int^{\bar{y}_{max}}_{-\bar{y}_{max}} d\bar{y} \;\text{e}^{-A}\rho_m(\bar{y})\rho_n(\bar{y})=\delta_{mn}$). With relation $dz=\text{e}^{(b-1)A}$dy, we get
\begin{eqnarray}\label{condtionforvector}
\int\rho_0^2 dz=\int\rho_0^2 \text{e}^{(b-1)A}dy \rightarrow
\int\text{e}^{\frac{-2v^2a}{9}(b+\sqrt{3b}\;\tau)\;y}dy ~~~\text{for}~~~y\rightarrow\infty,
\end{eqnarray}
which is in accord with that in Ref.~\cite{two-scalarbrane1}.

So if there is no coupling between the dilaton and the vector fields, the orthonormality condition for the zero mode becomes
\begin{eqnarray}
\int^{\bar{y}_{max}}_{-\bar{y}_{max}} d\bar{y} \;\text{e}^{-A}\rho_0^2(\bar{y})~~~~\propto~~~~\int^{\bar{y}_{max}}_{-\bar{y}_{max}} d\bar{y}~~~~~~<~~~~\infty,
\end{eqnarray}
which is the same with that in Ref.~\cite{BajcPLB2000}. However it is clear that the zero mode for the vector field can be localized on the brane, because the extra dimension $\bar{y}$ is finite.

While for the case that there is a coupling, it can be obtained from the relation (\ref{condtionforvector}) that the condition for the localization of the zero mode for the vector field is
$\tau\geq-\sqrt{b/3}$ for $0<b\leq1$ or $\tau>-1/\sqrt{3b}$ for $b>1$.

\begin{figure*}[htb]
\begin{center}
\includegraphics[width=5.5cm]{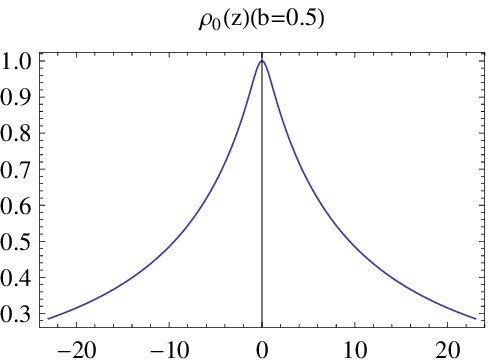}
\includegraphics[width=6.0cm]{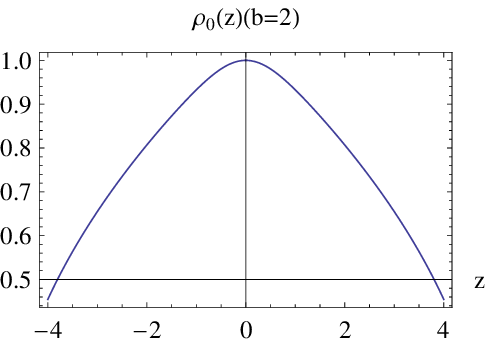}
\end{center}
 \caption{The shape of the zero mode for vector field $\rho_0(z)$ with $a=1,v=1,\tau=0$.}
 \label{figVzeromode}
\end{figure*}

Because the zero mode can be localized on the brane for different $b$, so there will no unstable KK modes. For $b<1$ there is no massive bound KK mode but some resonance may exist. And there will be finite number of the massive bound KK modes for $b=1$, but infinite ones for $b>1$.

\subsection{The Kalb-Ramond fields}

In this subsection we investigate the KR fields. The action of a KR field coupled with the dilaton is
\begin{equation}
S_{\text{KR}} = -\int d^{5}x \sqrt{-g}\; \text{e}^{\zeta \pi} H_{MNL}H^{MNL},\label{actionKRPhi}
\end{equation}
where $H_{MNL}=\partial_{[M}B_{NL]}$ is the field strength for the KR field, $H^{MNL}=g^{MO}g^{NP}g^{LQ}H_{OPQ}$ and $\zeta$ is the coupling constant. The equations of motion derived from this action and the conformal metric (\ref{Clinee}) read:
\begin{eqnarray}
 \text{e}^{\zeta\pi}\partial_\mu ( \sqrt{-g}H^{\mu\alpha\beta})
 +\partial_4(\sqrt{-g}\text{e}^{\zeta\pi}H^{4\alpha\beta})&=& 0, \\
 \text{e}^{\zeta\pi}\partial_\mu ( \sqrt{-g}H^{\mu4\beta})&=& 0.
\end{eqnarray}
If we choose the gauge $B_{\alpha4}=0$ and make a decomposition of the KR field
$B^{\alpha\beta}_{(n)}(x^\lambda,z)=\sum_n
\hat{b}^{\alpha\beta}_{(n)}(x^\lambda)U_{(n)}(z) \text{e}^{(-7-\sqrt{3}\zeta)A/2}$,
we will get the following Schr\"{o}dinger equation for the KK mode $U_{n}(z)$:
 \begin{eqnarray}
\big( -\partial^2_z+ V_{\text{KR}}(z)\big)U_n(z)
  =m_n^2 U_n(z),
  \label{SchEqK-R2}
\end{eqnarray}
where the effective potential $V(z)$ takes the following form
\begin{eqnarray}
V_{\text{KR}}=\frac{(1-\sqrt{3b}\zeta)^2}{4}A'^2
                         +\frac{\sqrt{3b}\zeta-1}{2}A''.
\label{VKRPi}
\end{eqnarray}
Provided the orthonormality condition $\int^{z_b}_{-z_b} dz
\;U_m(z)U_n(z)=\delta_{mn}$, the action of the KR field (\ref{actionKRPhi}) is reduced to
\begin{eqnarray}
 S_{\text{KR}} &=& -\sum_{n}\int d^4 x \sqrt{-\hat{g}}~
       \bigg( \hat{g}^{\mu'\mu}\hat{g}^{\alpha'\alpha}\hat{g}^{\beta'\beta}
       \hat{h}_{\mu'\alpha'\beta'}^{(n)}\hat{h}_{\mu\alpha\beta}^{(n)}
       +\frac{1}{3}m_n^2\hat{g}^{\alpha'\alpha}\hat{g}^{\beta'\beta}
       \hat{b}_{\alpha'\beta'}^{(n)}\hat{b}_{\alpha\beta}^{(n)}
       \bigg)
\label{actionKR2}
\end{eqnarray}
with $\hat{h}_{\mu\alpha\beta}^{(n)}=\partial_{[\mu}\hat{b}_{\alpha\beta]}$ the 4-dimensional field strength tensor.

\begin{figure*}[htb]
\begin{center}
\includegraphics[width=4.6cm]{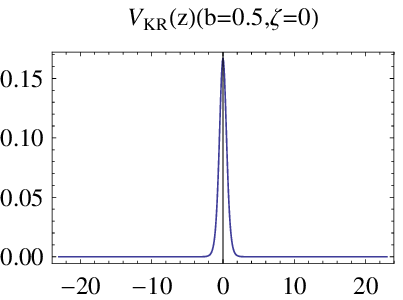}
\includegraphics[width=5.0cm]{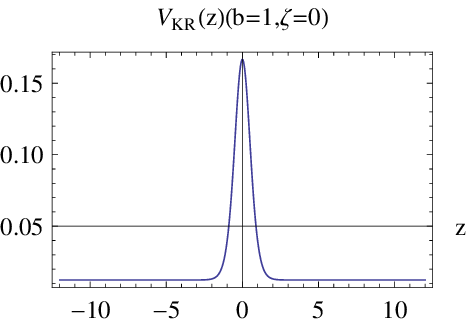}
\includegraphics[width=4.7cm]{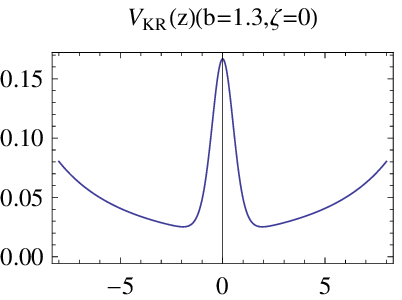}
\includegraphics[width=5.0cm]{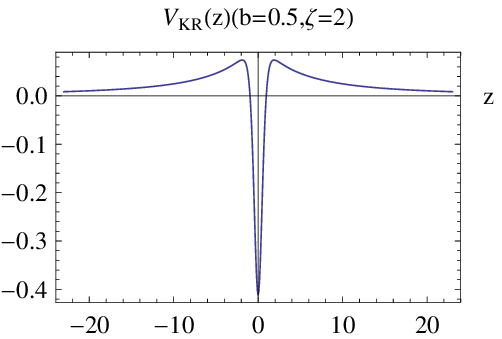}
\includegraphics[width=5.0cm]{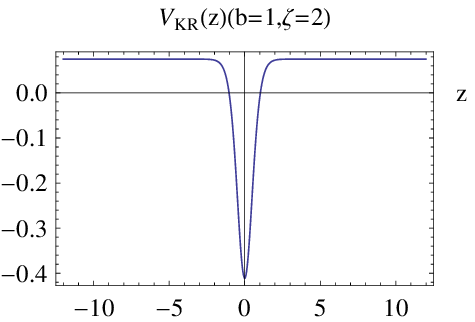}
\includegraphics[width=4.8cm]{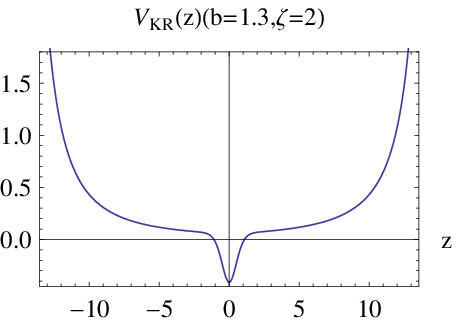}
\end{center}
 \caption{The shapes of the potentials of the KR field $V_{\text{KR}}(z)$. The parameters are set to $a=1, v=1$.}
 \label{figVKR}
\end{figure*}

The expression of the effective potential for the KR field $V_{\text{KR}}$ can be written as the function of $y$:
\begin{eqnarray}
V_{\text{KR}}=\text{e}^{2(1-b)A}\bigg[\frac{\sqrt{3b}\zeta-1}{2}\partial_y^2A
+\frac{(1-\sqrt{3b}\zeta)^2+2(1-\sqrt{3b}\zeta)(1-b)}{4}\big(\partial_yA\big)^2\bigg].
\end{eqnarray}
So with the braneworld solution (\ref{warpfactor}), the values of the potential
$V_{\text{KR}}(z)$ at $z=0$ and at the boundaries are
\begin{eqnarray}
V_{\text{KR}}(0)=-\frac{1}{6}a^2v^2(\sqrt{3b}\zeta-1)
\end{eqnarray} and
\begin{eqnarray}
V_{\text{KR}}(|z|\rightarrow z_b)=\frac{a^2v^4(\sqrt{3b}\zeta-1)
\big(\sqrt{3b}\zeta+1-2b\big)}{81(-\frac{2v^2a}{9}(b-1)z+1)^2}.
 \label{VKRzb}
\end{eqnarray}
In order to get negative value of the potential $V_{\text{KR}}(z)$ at $z=0$, the
parameter $\zeta$ should satisfy
\begin{eqnarray}
\zeta>1/\sqrt{3b}, \label{ConditionForZeta}
\end{eqnarray}
which is necessary for the localization of the zero mode. Therefore, it is clear that the zero mode of free KR fields ($\zeta=0$) can not be localized on the brane, and the coupling with the dilaton field $\pi$ is necessary for the purpose of localizing the KR field zero mode. From Eq. (\ref{VKRzb}), it can be seen that, under the condition (\ref{ConditionForZeta}), the potential is a volcano-like one
and a PT-like one for $0<b<1$ and $b=1$, respectively. While for $b>1$, the potential is divergent at the boundary $z=z_{max}$ with $\zeta\neq\frac{2b-1}{\sqrt{3b}}$, but vanishes with $\zeta=\frac{2b-1}{\sqrt{3b}}$. We plot the shapes of the
potential for KR field KK modes in the conformally flat space-time in
Fig.~\ref{figVKR}.

\begin{figure*}[htb]
\begin{center}
\includegraphics[width=5.2cm]{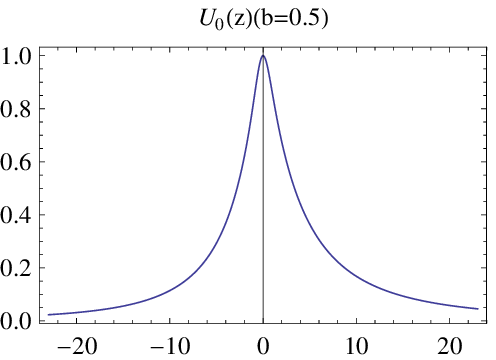}
\includegraphics[width=5.4cm]{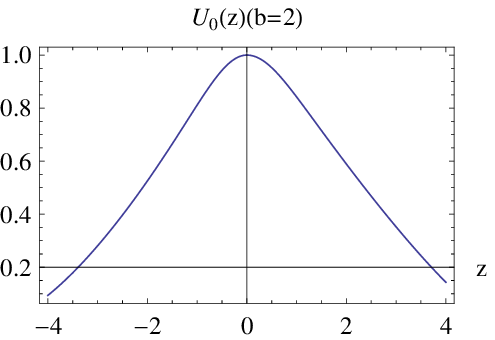}
\end{center}
 \caption{The shape of the zero mode for KR field $U_0(z)$ with $a=1, v=1,\zeta=2$.}
 \label{figKRzeromode}
\end{figure*}

So with the condition (\ref{ConditionForZeta}) we can obtain a zero mode for the KR
field by setting $m=0$:
\begin{eqnarray}
 U_0\varpropto \text{e}^{\frac{\sqrt{3b}\zeta-1}{2}A(z)}.
\end{eqnarray}
We also should check whether the zero mode for the KR fields can be localized on the brane through the orthonormality condition $\int^{z_b}_{-z_b} dz\;U_0^2(z)<\infty$. With the relation $dz=\text{e}^{(b-1)A}dy$ we get
\begin{eqnarray}
 \int U_0^2dz=\int U_0^2\text{e}^{(b-1)A}dy\rightarrow
  \int \text{e}^{-\frac{2av^2}{9}(\sqrt{3b}\zeta-2+b)y}dy
  ~~~~~~~~~~\text{for}~~~~y\rightarrow\infty.
\end{eqnarray}
Thus only when the coupling constant $\zeta$ satisfies $\zeta>1/\sqrt{3b}$ for $b\geq 1$ or $\zeta>\frac{2-b}{\sqrt{3b}}$ for $0<b<1$, the integral
$\int_{-z_b}^{z_b} U_0^2dz$ is finite, i.e., the zero mode for the KR field can be localized on the brane. The shape of the zero mode for KR field is plotted in Fig.~\ref{figKRzeromode}.

The spectrum structure of KR KK modes under the condition (\ref{ConditionForZeta}) is similar to the case of vector fields.

\subsection{The spin-1/2 fermion fields}
In the last subsection, we investigate the spin-1/2 fermion fields. Consider a massless spin $1/2$ fermion coupled with gravity and the background scalars $\phi$ and $\pi$ in 5-dimensional space, the Dirac action is
\begin{eqnarray}
 S_{1/2} &=& \int d^5 x \sqrt{-g} \bigg(\bar{\Psi} \Gamma^M (\partial_M + \omega_M)
 \Psi-\eta \bar{\Psi} F(\phi, \pi) \Psi\bigg)
 \label{DiracAction}
\end{eqnarray}
with $\eta$ the coupling constant and $F(\phi, \pi)$ the type of the
coupling. As in Ref.~\cite{Liu4,Liu5}, we have $\Gamma^M=(\text{e}^{-A}\gamma^{\mu},\text{e}^{-A}\gamma^5)$,
$\omega_\mu =\frac{1}{2}(\partial_{z}A) \gamma_\mu \gamma_5$,
$\omega_5 =0$, where $\gamma^{\mu}$ and $\gamma^5$ are the usual flat gamma matrices in the 4D Dirac representation. Then the 5-dimensional Dirac equation is read as
\begin{eqnarray}
 \left\{ \gamma^{\mu}(\partial_{\mu}+\hat{\omega}_\mu)
         + \gamma^5 \left(\partial_z  +2 \partial_{z} A \right)
         -\eta¡¡\text{e}^A F(\phi, \pi)
 \right \} \Psi =0, \label{DiracEq1}
\end{eqnarray}
where $\gamma^{\mu}(\partial_{\mu}+\hat{\omega}_\mu)$ is the 4-dimensional Dirac operator. Using the general chiral decomposition $ \Psi(x,z) = \text{e}^{-2A}\sum_n\big(\psi_{Ln}(x) f_{Ln}(z)
 +\psi _{Rn}(x) f_{Rn}(z)\big)$, we can get that $f_{Ln}(z)$ and $f_{Rn}(z)$ satisfy the following coupled equations
\begin{subequations}\label{CoupleEq1}
\begin{eqnarray}
 \left[\partial_z + \eta\;\text{e}^A F(\phi, \pi) \right]f_{Ln}(z)
  &=&  ~~m_n f_{Rn}(z), \label{CoupleEq1a}  \\
 \left[\partial_z- \eta\;\text{e}^A F(\phi, \pi) \right]f_{Rn}(z)
  &=&  -m_n f_{Ln}(z), \label{CoupleEq1b}
\end{eqnarray}
\end{subequations}
where $\psi_{Ln,Rn}(x)$ satisfy the four-dimensional massive Dirac equations
$\gamma^{\mu}(\partial_{\mu}+\hat{\omega}_\mu)\psi_{Ln}(x)=m_n\psi_{Rn}(x)$ and
$\gamma^{\mu}(\partial_{\mu}+\hat{\omega}_\mu)\psi_{Rn}(x)=m_n\psi_{Ln}(x)$. From the above coupled equations, we can get the following Schr\"{o}dinger-like equations for the KK modes of the left- and right-chiral fermions:
\begin{subequations}\label{SchEqFermion}
\begin{eqnarray}
  \big(-\partial^2_z + V_L(z) \big)f_{Ln}
            &=&m_n^2 f_{Ln},
   \label{SchEqLeftFermion}  \\
  \big(-\partial^2_z + V_R(z) \big)f_{Rn}
            &=&m_n^2 f_{Rn},
   \label{SchEqRightFermion}
\end{eqnarray}
\end{subequations}
where the effective potentials take the following forms
\begin{subequations}
\begin{eqnarray}\label{Vfermion}
  V_L(z)&=& \big(\eta\;\text{e}^{A}   F(\phi, \pi)\big)^2
     - \eta\partial_z \big(\;\text{e}^{A}   F(\phi,\pi)\big), \label{VL}\\
  V_R(z)&=&   V_L(z)|_{\eta \rightarrow -\eta}. \label{VR}
\end{eqnarray}
\end{subequations}
Moreover provided the following orthonormality conditions for
$f_{L{n}}$ and $f_{R{n}}$:
\begin{eqnarray}\label{Fermioncondition1}
 \int_{-z_b}^{z_b} f_{Lm}(z) f_{Ln}(z)dz
  &=&\delta_{mn},\\
\int_{-z_b}^{z_b} f_{Rm}(z) f_{Rn}(z)dz
  &=&\delta_{mn},\\
\int_{-z_b}^{z_b} f_{Lm}(z) f_{Rn}(z)dz&=&0,
\end{eqnarray}
we can obtain the standard 4-dimensional action for massive chiral fermions:
\begin{eqnarray}
 S_{1/2}=\sum_n\int d^4x \sqrt{-\hat{g}}\bar{\psi}_{n}(x)
 \big[\gamma^\mu(\partial_\mu+\hat\omega_\mu)-m_n\big]\psi_{n}(x).
\end{eqnarray}

From (\ref{VL}) and (\ref{VR}), it can be seen that, there must exist some kind of scalar-fermion coupling in order to localize the left- and right-chiral fermions. And if we demand that $V_{L,R}(z)$ are invariant under the reflection symmetry $z\rightarrow-z$, $F(\phi(z), \pi(z))$ should be an odd function of the extra dimension $y$. Thus we get $F(\phi(0), \pi(0))=0$ and $V_L(0)=-V_R(0)=-\eta\partial_z(F(\phi(0), \pi(0)))$, which results in that at most only one of the massless left- and right-chiral fermions could be localized on the brane. However, the masses of the massive KK modes of both chiral fermions are the same. In the following discussion, we only give the mass spectra for the left-chiral fermions.

In order to investigate the potentials of both chiral fermions, we use the relation $\partial_z=\text{e}^{(1-b)A}\partial_y$ to rewrite the potentials $V_{L,R}$ as the functions of $y$:
\begin{eqnarray}
V_L(y)&=& \eta\;\text{e}^{2A}\bigg[ \eta F(\phi, \pi)^2
     - \text{e}^{-bA}\bigg(\partial_y F(\phi,\pi)^2
         +F(\phi,\pi)\partial_y A\bigg)\bigg], \label{VL2}\\
V_R(y)&=&   V_L(y)|_{\eta \rightarrow -\eta}. \label{VR2}
\end{eqnarray}

Here we consider the case that the scalar-fermion coupling $F(\phi, \pi)$ takes the form $\text{e}^{\lambda \pi}\phi$ with $\lambda$ the dilaton-fermion coupling constant, then we can get the values of $V_{L,R}(z)$ at $z=0$ and $z\rightarrow z_b$ with the coordinate transformation $dz=\text{e}^{(b-1)A}dy$ and the expressions (\ref{VL2}) and (\ref{VR2}):
\begin{eqnarray}
 V_L(0)&=& -a v \eta,\\
 V_R(0)&=& a v \eta,\\
 V_L(z\rightarrow\pm z_b)&\rightarrow& v^2\eta^2
 [-\frac{2v^2a}{9}(b-1)z+1]^{\frac{2(1+\sqrt{3b}\lambda)}{b-1}}\nonumber\\
 &&+\frac{4av^3\eta}{9}(1+\sqrt{3b}\lambda)
 [-\frac{2v^2a}{9}(b-1)z+1]^{-\frac{(1+\sqrt{3b}\lambda)}{b-1}-1},\\\label{VLRInfinity}
 V_R(z\rightarrow\pm z_b)&=&V_L(z\rightarrow\pm z_b)|_{\eta \rightarrow -\eta}.
\end{eqnarray}
So we can see that for $1+\sqrt{3b}\lambda=0$, the values of the potentials at the boundaries will tend to a constant. And for $1+\sqrt{3b}\lambda>0$, the values of $V_{L,R}(|z|\rightarrow z_b)$ will always vanish for different $b$. While for $1+\sqrt{3b}\lambda<0$, it will always divergent. Thus there are three cases in total, which are decided by the relation between $\lambda$ and $b$.

\subsubsection{Case 1: $1+\sqrt{3b}\lambda=0$}
\begin{figure*}[htb]
\begin{center}
\includegraphics[width=6.0cm]{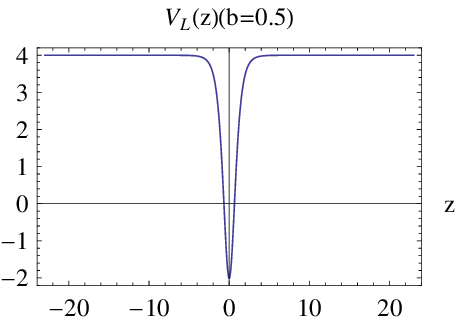}
\includegraphics[width=5.5cm]{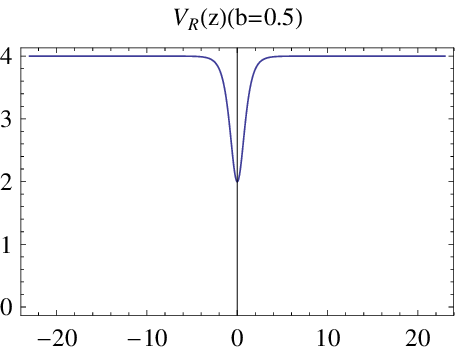}
\end{center}
 \caption{(Color online) The shapes of potentials $V_L(z)$ and $V_R(z)$ with $a=1,v=1,b=0.5,\eta=2$.}
 \label{figVL1}
\end{figure*}

For the case $1+\sqrt{3b}\lambda=0$, i.e., $\lambda=-1/\sqrt{3b}$, it can be seen that both $V_L(z \rightarrow\pm z_b)$ and $V_R(z \rightarrow\pm z_b)$ trend to a constant $v^2\eta^2$, which means the potentials are PT-like ones for left-chiral fermion with $\eta>0$ and for right-chiral fermion with $\eta>a/v$. So there exists a zero mode for left-chiral fermions for any positive value of $\eta$. And there will be a mass gap between the bound KK modes and continuous ones for both chiral fermions. The shapes of the potentials for a set of parameters are shown in Fig.~\ref{figVL1}.

From ({\ref{CoupleEq1a}}) we can solve the zero mode for the left-chiral fermions:
\begin{equation}
 f_{L0}(z) \propto \exp\left(-\eta\int^z_0 d\overline{z}
 \text{e}^{A(\overline{z})}F(\phi(\overline{z}), \pi(\overline{z}))\right).
  \label{zeroModefL0}
\end{equation}
In order to check whether the zero mode can be localized on the brane, we should check whether the following integral
\begin{eqnarray}
\int f_{L0}^2(z) dz
   && \propto\int \exp\left(-2\eta\int^z_0 d\overline{z}
 \;\text{e}^{A(\overline{z})}F(\phi(\overline{z}), \pi(\overline{y}))\right)dz\nonumber\\
   && = \int \exp\left(-2\eta\int^y_0 \text{e}^{(b+\lambda\sqrt{3b})A}d\overline{y}\right)\text{e}^{(b-1)A} dy
 \label{condition1}
\end{eqnarray}
is finite. With the relation $\lambda=-1/\sqrt{3b}$, we have
\begin{eqnarray}
\int f_{L0}^2(z) dz \rightarrow
     &&\int \text{e}^{-2\eta\;y}dy,
~~~~~~~~~~~~~~~~~~~~~~\text{for}~~b=1,\\
\int f_{L0}^2(z) dz\rightarrow
     &&\int
    \text{e}^{-2\frac{\eta}{k(b-1)}\text{e}^{k(b-1)y}}\text{e}^{k(b-1)y}dy,
~~\text{for}~~b\neq1,
\end{eqnarray}
with $k=-2v^2a/9$, which are both finite for different $b$. So the orthonormality conditions
(\ref{Fermioncondition1}) are always satisfied, namely, the zero mode for the left-chiral fermion can be localized on the brane with the dilaton-fermion coupling constant $\lambda=-1/\sqrt{3b}$.

Because the potentials are PT-like ones, so there exists mass gap for both chiral fermions with large $\eta>0$.

\subsubsection{Case 2: $1+\sqrt{3b}\lambda>0$}

\begin{figure*}[htb]
\begin{center}
\includegraphics[width=6cm]{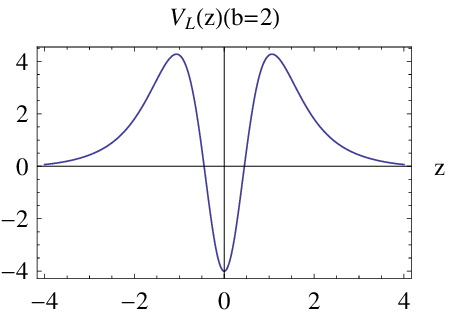}
\includegraphics[width=5.7cm]{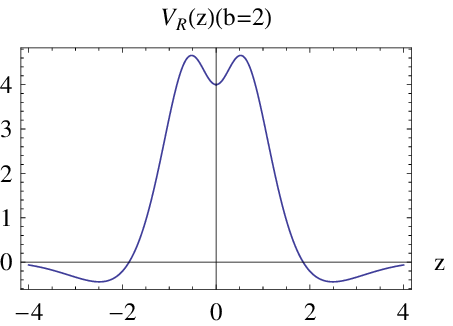}
\end{center}
 \caption{(Color online) The shapes of potentials $V_L(z)$ and $V_R(z)$. For the parameters are set to $a=1,v=1,b=0.5,\lambda=1,\eta=3$.}
 \label{figVL2}
\end{figure*}

For this case $1+\sqrt{3b}\lambda>0$, i.e., $\lambda>-1/\sqrt{3b}$, which includes $\lambda=0$, it can be seen that both $V_L(z \rightarrow \pm z_b)$ and $V_R(z \rightarrow \pm z_b)$ vanish, which means the potentials are volcano ones for left- and right-chiral fermions with large $\eta$. So there is only zero mode for left-chiral fermions for $\eta>0$. And there dose not exist a mass gap, but some resonances may appear for proper values of the parameters. The shapes of the potentials for both
chiral fermions are shown in Fig.~\ref{figVL2}.

Here we also need to check whether the zero mode (\ref{zeroModefL0}) satisfies the conditions (\ref{Fermioncondition1}), i.e., whether the zero mode can be localized on the brane. From (\ref{condition1}), we get that
\begin{eqnarray}
\int f_{L0}^2(z) dz\rightarrow &&\int
    \text{e}^{-\frac{2\eta}{k(b+\lambda\sqrt{3b})}
    \text{e}^{k(b+\lambda\sqrt{3b})y}}\text{e}^{k(b-1)y}dy.
    \label{condition2}
\end{eqnarray}
So we find that, for $\lambda > -1/\sqrt{3b}$ with $b>1$ or $-1/\sqrt{3b}<\lambda<-\sqrt{b/3}$ with $0<b<1$, the integral is finite, and the zero mode of the left-chiral fermion can be localized on the brane. But for $b=1$, the integral is infinite.

\subsubsection{Case 3: $1+\sqrt{3b}\lambda<0$}
\begin{figure*}[htb]
\begin{center}
\includegraphics[width=6cm]{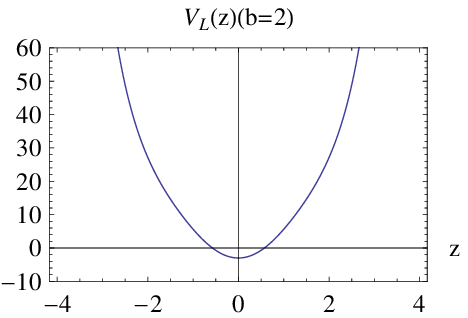}
\includegraphics[width=5.4cm]{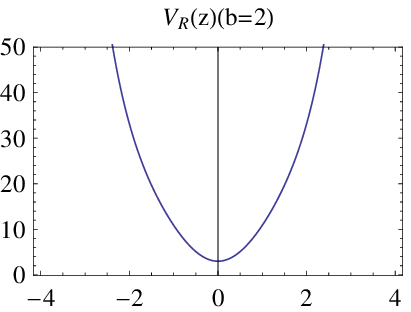}
\end{center}
\vskip -4mm \caption{(Color online) The shapes of potentials $V_L(z)$ and $V_R(z)$. The parameters are set to $a=1,v=1,b=1,\lambda=-1,\eta=3$.}
 \label{figVL3}
\end{figure*}

For the last case $1+\sqrt{3b}\lambda<0$, i.e., $\lambda<-1/\sqrt{3b}$, the potentials $V_{L,R}(z)$ are divergence at $z\rightarrow\pm z_b$. So there will be only bound KK modes for both chiral fermions, but there is only the zero mode for left-chiral fermion for positive value of $\eta$. We plot the shapes of the potentials for both chiral fermions in this case in Fig.~\ref{figVL3}.

From the relation (\ref{condition2}), we find that the integral $\int^{z_b}_{-z_b} f_{L0}^2 dz $ is always finite in this case, so the zero mode of left-chiral fermion (\ref{zeroModefL0}) can be localized on the brane.

\section{Discussions and conclusions}
\label{secConclusion}

In this paper, by presenting the shapes of the mass-independent potentials of KK modes in the corresponding Schr\"{o}dinger equations, we investigated the localization and mass spectrum of various bulk matter fields in a braneworld. The braneworld is generated by two interacting scalar fields, i.e., the kink $\phi$ and the dilaton $\pi$. There is a unique parameter $b$ in the solution, which leads to different distributions of the energy density of the system, and will effect the localization of various bulk matter fields.


For spin-1 vector fields coupled with the dilaton via
$\text{e}^{\tau\pi}F_{MN}F^{MN}$, when $\tau>-1/\sqrt{3b}$,
the effective potential is a volcano-like potential for $0<b<1$
and a PT-like one for $b=1$. But it will diverge at the boundaries of the extra dimension for $b>1$ but $\tau\neq\frac{2b-3}{\sqrt{3b}}$, and vanish for $\tau\neq\frac{2b-3}{\sqrt{3b}}$.
And there is always a localized vector zero mode
for $\tau>-\sqrt{b/3}$ with $0<b\leq1$, or $\tau>-1/\sqrt{3b}$ with $b>1$. So when
$\tau=0$, i.e., the case without coupling with the dilaton, the vector zero mode can
also be localized on the brane for any $b>0$.

For KR fields coupled with the dilaton $\pi$, using the conformal metric and the KK decompositions, we also
obtained the Schr\"{o}dinger equations for the KR KK modes.
When the coupling constant $\zeta$ satisfies $\zeta>1/\sqrt{3b}$ with $b\geq1$, or
$\zeta>\frac{2-b}{\sqrt{3b}}$ with $0<b<1$, there will be a localized zero mode. And
with $\zeta>1/\sqrt{3b}$, there are also three types of potentials for different $b$,
which is similar to the vector fields.

While for spin $1/2$ fermion fields, in order to localize the left-chiral and right-chiral fermions on the brane, some kind of coupling between the fermions and the background scalars should be introduced. In the paper, we considered the coupling type $\eta\bar{\Psi}\text{e}^{\lambda\pi}\phi\Psi$ with positive Yukawa coupling constant $\eta$. We found that the relation between the constant $\lambda$ and $b$ is crucial. With different relations, there will be three types of potentials.

If $\lambda=-1/\sqrt{3b}$, the potentials for left- and right-chiral fermions with lager $\eta$ are PT-like ones. So there is a zero mode for left-chiral fermion and the zero mode can be localized on the brane for different $b$. The number of the massive bound KK modes for both chiral fermions is finite for $b>0$.

For $\lambda>-1/\sqrt{3b}$, the potentials for both chiral fermions become volcano potentials. Therefore, there is no massive bound KK mode for this case. The zero mode for left-chiral fermion can be localized on the brane for $b>1$ and $\lambda>-1/\sqrt{3b}$, or $b<1$ and $-1/\sqrt{3b}<\lambda<-\sqrt{b/3}$. However, for $b=1$, the zero mode for left-chiral fermion can not be localized on the brane.

The potentials for left- and right-chiral fermions have infinite potential wells when $\lambda<-1/\sqrt{3b}$, so there will be only bound KK modes for any $b>0$. The zero mode for the left-chiral fermion can be localized on the brane.

\section{Acknowledgement}

This work was supported by the Program for New Century Excellent
Talents in University, the Huo Ying-Dong Education Foundation of
Chinese Ministry of Education (No. 121106), the National Natural
Science Foundation of China (No. 11075065), the Doctoral Program Foundation of
Institutions of Higher Education of China (No.
20090211110028), and the Natural Science Foundation of Gansu Province,
China (No. 096RJZA055).

\end{document}